\documentclass{aa}
\usepackage{graphicx}
\usepackage{txfonts}
\begin{document}
   \title{An analysis of the durations of Swift Gamma-Ray Bursts}


   \author{Zhi-Bin Zhang
          \inst{1, 2}
          \and
          Chul-Sung Choi
          \inst{1}
          }
   \offprints{Z. B. Zhang}

   \institute{International Center for Astrophysics, Korea
Astronomy and Space Science Institute, 36-1 Hwaam, Yusong, Daejon
305-348, South Korea; \email{zbzhang@kasi.re.kr}
         \and
             Yunnan Observatory, National Astronomical Observatories, Chinese Academy of
        Sciences,
            P. O. Box 110, Kunming 650011, China\\
             }



  \abstract
   {Swift detectors are found to be more sensitive to long-soft bursts than pre-Swift missions. This
   may largely bias the distribution of durations and then classification of gamma-ray bursts. }
   {In this paper, we systematically investigate the duration distribution of gamma-ray bursts in the Swift era via comparison
   with that of pre-Swift bursts.}
   {For the purpose of this study, statistical methods such as the K-S test and
   linear/non-linear fitting analysis have been used to examine the duration properties of Swift bursts in both observer and source frames.}
   {For 95 GRBs with known redshift, we show that two-lognormal distributions of duration are
clearly divided at $T_{90}\simeq2$~s. The intrinsic durations also
show a bimodal distribution but shift systematically toward the
smaller value and the distribution exhibits a narrower width
compared with the observed one. Swift long bursts exhibit a wider
duration dynamic range in both observer and source frame in
comparison to pre-Swift long bursts.}
   {We find that Swift bursts and pre-Swift ones can share the same criterion of classification in terms of duration at 2 seconds,
   although both monitors have large difference with respect to sensitivity of a given energy band.}

\titlerunning{Properties of GRB duration in the Swift era}
\authorrunning{Z. B. Zhang \& C. S. Choi}
   \maketitle
%

\section{Introduction}

   Cosmic gamma-ray bursts (GRBs) are the most violent explosions
occurring at cosmological distances in the universe. When a GRB
takes place, satellites can monitor its temporal variability in the
$\gamma$-ray energy band. The duration of the burst, $T_{90}$, is
defined as the time interval in which the integrated photon counts
increase from 5\% to 95\% of the total counts. Based on an analysis
of durations using initial BASTE data, Kouveliotou et al. (1993)
divided GRBs into two classes, i.e., long GRBs (LGRBs) with $T_{90}
> 2$~s and short GRBs (SGRBs) with $T_{90}<2$~s. The dichotomy has
been justified by subsequent investigations (e.g., Mao, Narayan \&
Piran 1994; Katz \& Canel 1996; Meegan et al.1996; Paciesas et al.
1999; Fishman 1999). In fact, the best parameters of a two-component
lognormal fit to the distribution data were first obtained by
McBreen et al. (1994). This fit has been supported by the current
BATSE data with peak flux information (Horv\'{a}th 2002; Nakar
2007).

So far, much evidence showing the difference between two classes has
been discovered and presented (see Zhang 2006 for review). For
example, the spectra of LGRBs are softer than that of SGRBs in
general. Besides, the pulse profiles of SGRBs are on average more
symmetric than those of LGRBs (Zhang \& Xie 2007). The current Swift
observations show that LGRBs have their median cosmological redshift
$z_{m}\sim 2.0$ higher than that of SGRBs $z_{m}\sim 0.4$. All these
differences suggest that both LGRBs and SGRBs, most likely, are
distinct physical phenomena and produced due to model-independent
emission engines (e.g., Bal\'{a}zs et al. 2003). However, it is not
clear what causes these differences, especially the origin of
bimodal $T_{90}$ distribution.

Koshut et al. (1996) pointed out that the observed duration
distribution may vary with instruments. It is therefore necessary to
investigate if there exists a new GRB class and/or what physical
factors produce such properties (Gehrels et al. 2004). Naturally, we
focus our study on the related issues of GRB classification using
the updated Swift
data\footnote[1]{http://swift.gsfc.nasa.gov/docs/swift/archive/grb\_table.html}$^,$
\footnote[2] {http://grad40.as.utexas.edu/tour.php (\textit{GRBlog}
site)}$^,$ \footnote[3]{http://www.mpe.mpg.de/~jcg/grbgen.html}.

\section{Observation and data selection}

The higher sensitivity and angular resolution of Swift make it
superior to previous space telescopes (e.g. BATSE, BeppoSAX and
HETE-2), accurate follow-up observations have further added to its
advantage (Gehrels et al. 2004). It can detect on an average about 2
GRBs per week within a 2 sr field of view, which is about two times
more than that of pre-Swift missions (M\'{e}sz\'{a}ros 2006).
However, the detection rate of SGRBs to total GRBs is much lower
($\sim 8\%$) than the rate by BATSE ($\sim 18 \%$), which is
attributed to both their different energy responses and the
relatively high spectral hardness of SGRBs (e.g., Band 2006a, b;
Gehrels et al. 2007). On the other hand, the lower sensitivity to
short duration bursts of Swift, relative to BATSE, makes it
accumulate relatively lower counts, comparable with the number of
background counts (Band 2006a). The effect of the instrument may
cause the detection rate of SGRBs to be somewhat underestimated.

In order to study the intrinsic properties of GRBs, we selected six
data sets, namely s1 - s6, as listed in Table 1. As of 2007 July 1,
Swift has detected 75 LGRBs (s1) and 20 SGRBs (s2) with known
duration and redshift. The $E_p$ in the observed $\nu F_{\nu}$
energy spectra are chosen to characterize the spectral hardness
relations with duration. Here, 44 GRBs from s1 have also the
available $E_p$ values and constitute our sample s3. Out of the s2,
only 11 sources have the measured $E_p$ and are employed to build
the sample s4. Unfortunately, the remaining 9 bursts in s2 (i.e.,
\object{GRB 050202}, \object{GRB 050906}, \object{GRB 050925},
\object{GRB 051105A}, \object{GRB 051210}, \object{GRB 060313},
\object{GRB 070209}, \object{GRB 070406} and \object{GRB 070429B})
do not have the measured redshifts. For these we assigned a redshift
value of $z$=0.5 to the 9 bursts, approaching the median redshift of
$z$=0.4, as assumed by Norris \& Bonnell (2006). These sources were
included in the present study to improve the statistics. However, we
found that the choice between $z$=0.5 and $z$=0.4 contributes only a
small relative error of $\sim$ 0.07, implying that the final results
are not sensitive to the above assumed redshift values. In our fifth
sample set, s5, we include 48 pre-Swift LGRBs whose $z$ and $T_{90}$
are already measured, in which 18 sources, less than half ($\sim$
38\%) of the 48 pre-Swift bursts, are detected by the BATSE mission
and constitute our sample s6.

\begin{table} \centering \caption{GRB samples of known durations ($T_{90}$),
redshifts ($z$) and/or peak energies ($E_{p}$)} \label{Tab.1
samples}
\begin{tabular}{c c clr}
\hline
Sample & No.&Class & Parameters &Ref. \\
 \hline
s1$^{\dag}$& 75&LGRBs&$T_{90}$; $z$&1; 1; -\\
s2$^{\dag}$& 20&SGRBs&$T_{90}$; $z$&1; 1; - \\
s3$^{\dag}$& 44&LGRBs&$T_{90}$; $z$; $E_p$&1; 1; 4 \\
s4$^{\dag}$& 11&SGRBs&$T_{90}$; $z$; $E_p$&1; 1; 4 \\
s5$^{\ddag}$& 48&LGRBs&$T_{90}$; $z$&2, 3; 5; -\\
s6$^{\sharp}$& 18&LGRBs&$T_{90}$; $z$&2, 3; 5; -\\
\hline
\end{tabular}
\begin{flushleft}
\textit{Note}: The references are given in order for $T_{90}$, $z$
and $E_p$. Data sets: $^{\dag}$ Swift; $^{\ddag}$ pre-Swift;
$^{\sharp}$ BATSE. Of the 20 SGRBs, \object{GRB 050709}, is
adopted from HETE-2.\\

\textit{Reference}: 1. Swift public webpage; 2. Ghirlanda,
Ghisellini \& Lazzati 2004; 3. \textit{GRBlog} website; 4. Butler et
al. 2007; 5. Friedman \& Bloom 2005.
\end{flushleft}
\end{table}


\section{results}


\subsection{Distributions of durations in the Swift era}


To check whether the distributions of $T_{90}$ are significantly
different, indicative of a dependence on the on-board instruments,
we first obtained the distributions of Swift bursts in both observer
and source frames. Then, we compare the durations of LGRBs between
Swift and pre-Swift/BATSE missions.

\subsubsection{Observed $T_{90}$ distribution}

The accuracy of $T_{90}$ measurements is in principle affected by
several factors including the identification of the time interval of
a burst, the sensitivity of instrument, background modeling, the
time resolution of the data, and the detailed shape of the burst
time profile, etc. Figure 1 shows the $T_{90}$ distribution for the
95 Swift GRBs, which include s1 and s2 samples (see Table 1). The
best fit with a two-lognormal function gives the center values
($T_{90, p1}=0.28_{-0.09}^{+0.14}$~s and $T_{90,
p2}=42.83_{-4.45}^{+4.60}$~s) and the widths
($w_{1}=19.05_{-11.11}^{+24.60}$~s and
$w_{2}=18.20_{-3.41}^{+4.19}$~s) with the reduced Chi-square
$\chi^{2}$/dof = 0.67, which are roughly consistent with those
calculated from the BATSE data (McBreen et al. 1994; Meegan et al.
1996; Paciesas et al. 1999; Horv\'{a}th 2002; Nakar 2007). Note that
the number of objects classified as belonging to the two lognormals
to itself has been allowed to be a free parameter. The superposed
function has a minimum around 2 s as found by Kouveliotou et al.
(1993), indicating that the Swift sources are also divided into two
classes, SGRBs and LGRBs, although the Swift is more sensitive to
long soft bursts than the BATSE (Band 2006a,b; Gehrels et al. 2007).
This is an interesting result since the different detectors with
diverse bandwidth sensitivity do not indeed affect the
classification of GRBs in terms of duration.

\begin{figure}
\includegraphics[width=8cm]{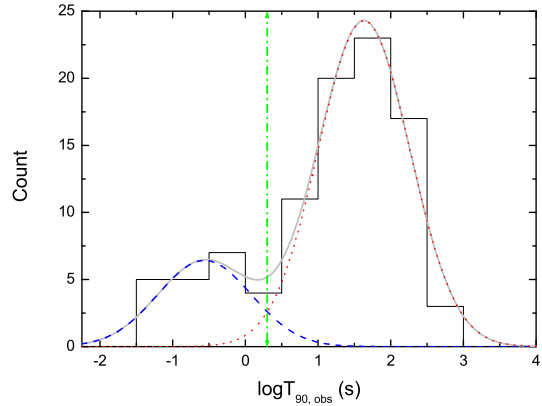}
  \caption{Bimodal distribution of durations for the 95 GRBs (s1 and s2; histogram)
and two-component lognormal fit to the data (solid line). The GRBs
are divided into two classes at $T_{90}\simeq2$ s (vertical line):
LGRBs (dotted line) and SGRBs (dashed line).}
  \label{Fig1 T90 dis}
\end{figure}
\subsubsection{Intrinsic $T_{90}$ distribution}

One of the great Swift progresses is the increase in number of
higher redshift sources. It is known that the median redshift of
Swift LGRBs, $z_{m}\sim2$, is roughly two times larger than that of
pre-Swift. As pointed out by previous authors (Bagoly et al. 2006;
Jakobsson et al. 2006), the difference of redshift distributions
between the two samples is statistically significant. This may lead
to an evident discrepancy between the two intrinsic $T_{90}$
distributions. We therefore utilize the Swift sources to explore
this possibility.

The potential spectrum evolution, as a result of cosmological
redshift, can usually cause high energy $\gamma$ photons to shift
into or out of the finitely sensitive bandwidth of a given detector.
Note that here we had neglected the effect of a burst's spectrum
itself softening with time (Ford et al. 1995). In this case, the
transformation of $T_{90}$ from observer frame to source frame is
generally expressed as $T_{90, int}$=$T_{90}/(1+z)^{\omega}$, in
which $\omega=0.6$ or 1, corresponding to energy stretching or not
(Fenimore \& Bloom 1995; M\'{e}sz\'{a}ros \& M\'{e}sz\'{a}ros 1995;
M\'{e}sz\'{a}ros \& M\'{e}sz\'{a}ros 1996; Horv\'{a}th,
M\'{e}sz\'{a}ros \& M\'{e}sz\'{a}ros 1996; Bal\'{a}zs et al. 2003).
Here, we consider a simple case of $\omega=1$ throughout this work,
which the intrinsic duration is given by $T_{90,
int}$=$T_{90}/(1+z)$. We calculate the intrinsic duration ($T_{90,
int}$) distribution for the 95 GRBs and compare it with the observed
one. As shown in Figure 2, the $T_{90, int}$ has a bimodal
distribution and is significantly shifted toward shorter durations
than the observed one. The best fit with a two-lognormal function
gives two centers ($T_{90, p1}=0.13_{-0.05}^{+0.12}$~s, $T_{90,
p2}=12.30_{-1.83}^{+2.15}$~s) and two widths
($w_{1}=10.96_{-8.14}^{+31.69}$~s and
$w_{2}=17.38_{-5.63}^{+8.32}$~s) with $\chi^{2}$/dof = 0.92,
indicating that the distribution of $T_{90, int}$ is indeed bimodal
but systematically narrower and shifted towards low value of
durations in comparison to the observed one. A Kolmogorov-Smirnov
(K-S) test returns the statistic $D=0.27$ with a probability of
$P=0.001$, suggesting the intrinsic and observed duration
distributions are drawn from the different parent populations. This
result is well consistent with the theoretical prediction by Bromm
\& Loeb (2002), where they assumed the formation of all GRBs tightly
follows the cosmic star formation history. However, some
instrument-dependent factors such as the selection effects and
discrepancy between different detectors can distort the resultant
parent distribution of $T_{90, int}$ in a sense. Once the
disturbances are really reduced from the observed distribution, the
remainder is the true parent distribution correlated with some
physical predictions. In this case, our current result can offer
some corroborative statistical evidence that the rate of GRBs may
really trace the star formation history, partly because the
redshifts of SGRBs had not been measured before Swift.

\begin{figure}
\includegraphics[width=8cm]{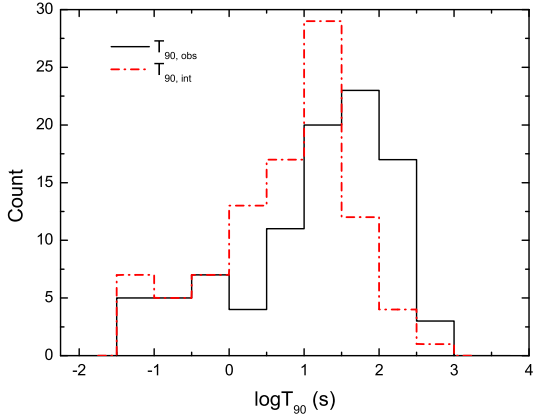}
  \caption{Comparison between the observed (solid line the same as Fig. 1) and the
intrinsic (dashed-dotted line) $T_{90}$ distributions.}
  \label{Fig2 T90-in dis}
\end{figure}
%
\subsubsection{$T_{90}$ of LGRBs: Swift vs. pre-Swift}
\begin{figure}
\resizebox{8cm}{12cm}{\includegraphics{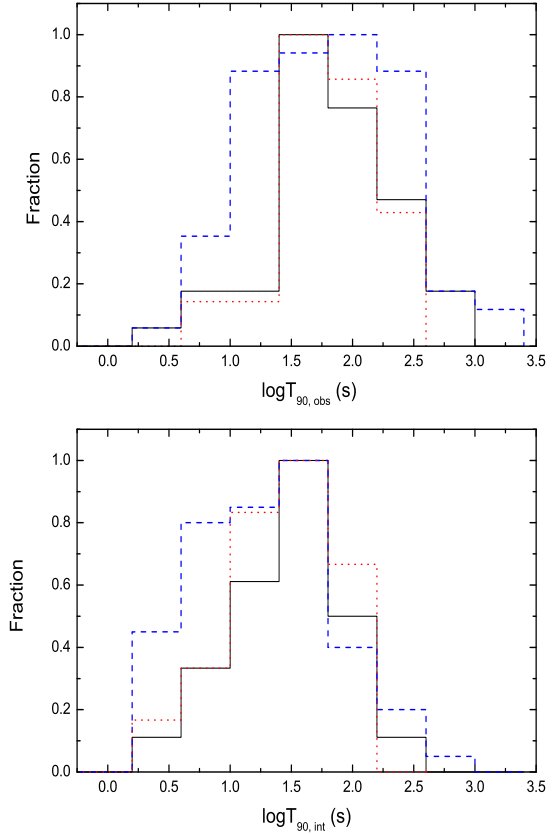}}
  \caption{The normalized duration distributions for the Swift (s1;
dashed line), pre-Swift (s5; solid line) and BATSE (s6; dotted line)
LGRBs in the observer (\textit{upper} panel) and the source
(\textit{lower} panel) frames.}
  \label{Fig3 T90 comp}
\end{figure}
$T_{90}$ may be wrongly estimated because of the factors pointed out
in section 3.1.1. In particular, the measured duration has been
found to be instrument-dependent (e.g., Koshut et al. 1996). We
examined the dependence of $T_{90}$ on the instruments by
contrasting the normalized distributions of different detectors in
Figure 3. For this comparison, three samples including s1, s5, and
s6 are applied.

A lognormal fit to the six distributions gives the best parameters
listed in Table 2. We find the Swift sample obviously exhibits a
wider duration distribution than the pre-Swift and BATSE did in both
observer and source frames, while the widths of pre-Swift and BATSE
distributions of duration are robustly equal. This implies that
Swift can detect GRBs in a wider dynamic range of $T_{90}$, i.e.,
the fraction of longer and shorter Swift LGRBs are significantly
higher than that of pre-Swift/BATSE LGRBs, except the increased
detection rate of LGRBs due to the higher sensitivity of BAT loaded
on Swift to long soft bursts (Band 2006 a,b). Considering the fact
that Swift bursts can still be separated at $T_{90}=2$ s, we predict
that the width of $T_{90}$ distribution of Swift SGRBs could be
equally wider in comparison to pre-Swift detectors. Furthermore, we
see from Table 2 that the distribution centers of $T_{90, obs}$ are
very close to each other, while for the $T_{90, int}$ distribution,
the centers of the pre-Swift and BATSE bursts are much more close
but significantly larger than that of Swift bursts. It is
interesting to note that the K-S test to the pre-Swift and BATSE
bursts provides a very large probability of $P\sim0.99$ in both
observer and source frames showing that the s5 and s6 samples are
consistent with being drawn from the same parent distribution.
As for the Swift $T_{90, int}$ distribution, the obvious decrease of
the center value is attributable to the relative increase of the
fraction of high redshift sources, which could be caused by the
ubiquitous threshold effect of different instruments (e.g., Bromm \&
Loeb 2002; Band 2006a,b).

\begin{table*}
\centering \caption{The best-fit parameters to the three data sets}
\label{Tab.2 for fig 3}
\begin{tabular}{ccccccc}
\hline
&  \multicolumn{3}{c}{log$T_{90, obs}$}& \multicolumn{3}{c}{log$T_{90, int}$}\\
\hline
Sample &$\mu$&$w$&$\chi^{2}/dof$&$\mu$&$w$&$\chi^{2}/dof$ \\
 \hline
s1  &1.57$\pm0.04$&1.31$\pm$0.08&2.4 &1.06$\pm$0.04&1.23$\pm$0.09&2.8\\
s5  &1.63$\pm$0.04&0.90$\pm$0.09&2.5 &1.32$\pm$0.03&0.89$\pm$0.05&0.9\\
s6  &1.62$\pm$0.04&0.76$\pm$0.07& 0.5&1.29$\pm$0.03&0.92$\pm$0.07&0.3\\
\hline
\end{tabular}
\begin{flushleft}
\textit{Note}: The fitted parameters for the distribution center
($\mu$) and the width ($w$) are given in a logarithmic scale.
\end{flushleft}
\end{table*}

\subsection{Spectral hardness relation}

\begin{figure*}
\resizebox{18cm}{6cm}{\includegraphics{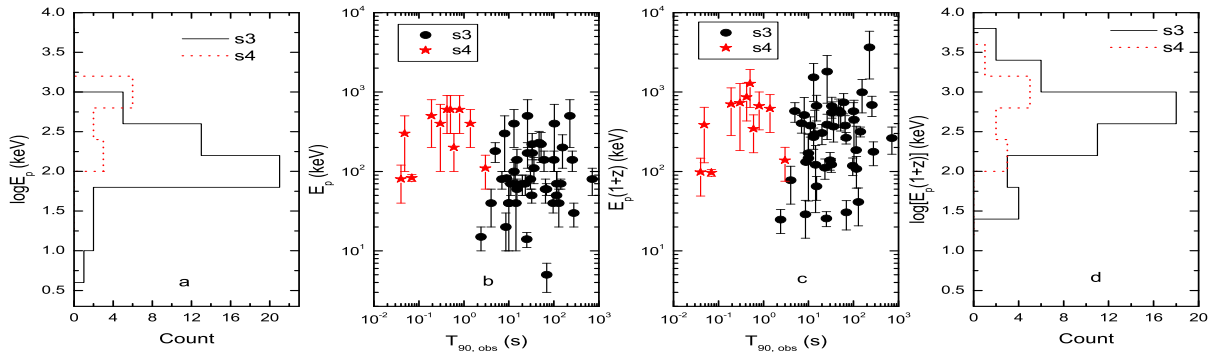}}
\caption{Relations of $T_{90}$ with $E_p$ (panel \textit{b}) and
$E_{p, i}$ (panel \textit{c}). Panels \textit{a} and \textit{d} show
the distributions of the $E_p$ and $E_{p, i}$ in logarithmic
timescale, respectively.}
  \label{Fig4 hardness}
\end{figure*}
In theory, the peak energy $E_p$ can be used to reflect the photon
components, similar to hardness ratio between different energy
channels. The intrinsic peak energy ($E_{p,i}$) in the source frame
is related with the $E_{p}$ in the observer frame by $E_{p, i}=E_p
(1+z)$. To verify the relation of spectral hardness with duration,
we made plots of peak energy versus $T_{90, obs}$ for the samples s3
and s4 as shown in Figure 4\textit{b} and 4\textit{c}.
A linear correlation analysis gives the coefficient of $r=-0.27$
with probability $P=0.07$ for Figure 4\textit{b} and $r=0.004$ with
$P=0.98$ for Figure 4\textit{c}, indicating that the tendency of
SGRBs relatively harder than LGRBs exists in the observer frame and
almost disappears in the source frame. The former accords with the
previous findings, for example, using the BATSE data (e.g.,
Kouveliotou et al. 1993) or the Swift data (e.g., Band 2006a).

\textit{We also compare the peak energy of LGRBs and SGRBs in both
observer (Figure 4\textit{a}) and source frame (Figure
4\textit{d}).} It needs to point out that more cosmological redshift
correction is required to give LGRBs when we convert the physical
variables from the observer frame to the source frame. As a result,
the trend of the relative spectral softness for LGRBs (or the
relative hardness for SGRBs) weakens more or less. For our samples,
the long and short bursts have the median values of
$74^{+120}_{-45}$ keV and $398^{+493}_{-220}$ keV, respectively, in
the observer frame, and $302^{+631}_{-202}$ keV and
$617^{+896}_{-365}$ keV, respectively, in the source frame. A K-S
test returns the different probabilities of $P=0.01$ for Figure
4\textit{a} and $P=0.15$ for Figure 4\textit{d}, which hints that
the $E_{p, i}$ of long and short bursts may be drawn from the same
parent population.

\section{Discussion}

The duration distribution from the third BATSE catalog is suggested
to have a three-component lognormal form (Horv\'{a}th 1998;
Mukherjee 1998; Hakkila et al. 2000). Similarly, the duration
distribution of the current 4B
catalog\footnote[4]{http://www.batse.msfc.nasa.gov/batse/grb/duration/}
exhibits the possible existence of an intermediate group.
Further studies show that the third group is either the excess of
SGRBs with low fluence (Hakkila et al. 2003) or the softest LGRBs
(Horv\'{a}th et al. 2006). It is still controversial whether there
exists the third class or not, due to the lack of a physical
explanation (Horv\'{a}th et al. 2006; see however, Chattopadhyay et
al. 2007). The third group might be caused by an instrumental bias
which reduces the durations of faint LGRBs (e.g., Hakkila et al.
2000; Hakkila et al. 2003), suggesting their presence \textbf{is}
not physical but phenomenological. Therefore, the bimodality of the
duration distribution is widely accepted by most people today.

Basically, what causes the bimodality is an interesting but unsolved
problem owing to the absence of direct observational evidence,
although several works have tried to explain the bimodality within
different scenarios. It was suggested that the different spin axes
of millisecond pulsars can interpret the two GRB classes (Usov 1992;
Yi \& Blackman 1998). Subsequently, Huang et al. (2003) studied the
neutron star kick model (Dar \& Plaga 1999) in detail and proved
that this model can successfully account for the two-lognormal
distributions if the central engine has a neutron star of high kick
velocity larger than $\sim1000$ km $s^{-1}$. Yamazaki, Ioka \&
Nakamura (2004) put forward the so-called unified model consisting
of multiple subjets within an inhomogeneous main jet. Using this
model, Toma, Yamazaki \& Nakamura (2005) explained that the
bimodality originates from discrete emitters in the main jet. They
also predicted that the two kinds of bursts should have the same
origin, i.e., supernovae, instead of the leading models, which
predict that the LGRBs and SGRBs are produced due to the core
collapse of a massive star and the merger of double compact objects,
respectively (see e.g., Cheng \& Lu 2001; Zhang \& M\'{e}sz\'{a}ros
2004; Piran 2005; M\'{e}sz\'{a}ros 2006 and Lee \& Ramirez-Ruiz 2007
for reviews). However, further investigations on the unified model
showed that the bimodal distribution could be reproduced only for
some special parameters (Janiuk et al. 2006).

Recently, from an independent analysis of distinct timescales, Zhang
et al. (2007) suggested that LGRBs occur at larger distances from
the central engine while SGRBs at smaller distances, i.e., two
distinct $\gamma$-ray emission regions may result in two different
properties of GRBs, including the varieties of pulse profiles, as
mentioned above. In this study, they also pointed out the fact that
LGRBs usually have long positive spectral lag (e.g., Norris, Marani
\& Bonnell 2000; Daigne \& Mochkovitch 2003; Chen et al. 2005) and
SGRBs have negligible lag (Norris \& Bonnell 2006; Zhang et al.
2006) can be naturally explained under the assumption that the
curvature effect is a main contributor to the spectral lags (Ryde
2005; Shen, Song \& Li 2005). The predication of distinct emission
regions for different GRB classes and/or the nature of the
bimodality still need to be clarified with more accurate
observations in the future.

\section{Conclusions}
  \begin{itemize}
\item We find that the observed
durations of Swift GRBs have two-lognormal distributions divided
clearly at $T_{90}\simeq2$ s. This implies that the classification
in terms of duration is unchanged from pre-Swift to Swift era.
\item The intrinsic durations also show a bimodal distribution but
shifted systematically toward the smaller value and the distribution
exhibits a narrower width relative to the observed distribution.
\item The $T_{90, int}$ distributions of Swift and pre-Swift/BATSE
LGRBs are significantly different in the width and center values.
\item Swift bursts have a wider duration dynamic range than
pre-Swift and BATSE bursts.
\item The spectra of the SGRBs are predominantly harder than the
LGRBs, confirming the previous results from the BATSE bursts.
However, the trend of LGRBs with relatively softer spectrum largely
weakens in the source frame.
\end{itemize}
%
%


%

\begin{acknowledgements}
We thank David L. Band and Attila M\'{e}sz\'{a}ros for their helpful
comments and suggestions. The authors would like to thank the
anonymous referee for a thorough and constructive report which has
led to a substantial improvement of the paper. We thank Istvan
Horv\'{a}th for good communication and G. Maheswar for critical
reading of the manuscript. Z. B. Z wishes to acknowledge the
postdoctoral fellowship in Korea Astronomy and Space Science
Institute (KASI).
\end{acknowledgements}

\end{document}